\documentstyle[prl,aps,epsf,floats]{revtex}
\begin{document}

\draft

%% --- TWO COLUMN STYLE : uncomment following 2 lines
\twocolumn[\hsize\textwidth\columnwidth\hsize\csname@twocolumnfalse%
\endcsname
%% ---------------------------

\title{Random Matrix Model for Superconductors in a Magnetic Field}
\author{Safi R. Bahcall}
\address{Department of Physics, University of California, 
Berkeley CA 94720}
\maketitle

\begin{abstract}
We introduce a random matrix ensemble for bulk type-II superconductors
in the mixed state and determine the single-particle excitation
spectrum using random matrix theory.  The results are compared
with planar tunnel junction experiments in PbBi/Ge thin films.  More low
energy states appear than in the Abrikosov-Gor'kov-Maki or
Ginzburg-Landau descriptions, consistent with observations.
%% \vspace*{.1in}
\end{abstract}

%% 74.60.-w : type II superconductivity
%% 74.25.Jb : ... and electronic structure
%% 74.50.+r : ... tunneling ...
%% 05.40.+j : random processes

%% --- TWO COLUMN STYLE : uncomment following lines
 ]
 \def\bottomfraction{.9}
 \def\textfraction{.1}
%% ---------------------------
%% (above \def's make the figure come out properly located)

Random matrices have been used to understand the distribution of level
spacings and widths in nuclei \cite{wigner,morenuclei}, complex atoms
\cite{morenuclei}, small metallic particles \cite{gorkov,altschuler},
and quantum systems whose classical limits are chaotic
\cite{quanchaos}.  Random matrix models have also appeared in the
context of solving certain SU(N)-invariant field theories in the large
N limit \cite{thooft,BIPZ} and discretizing two dimensional quantum
gravity \cite{ginsparg}.  In this paper we consider a different
context: using a random matrix model to solve the electronic structure
problem posed by the BCS description of a superconductor in a magnetic
field.

The BCS description for an ideal electronic system, and the more
general description for a realistic system, can be formulated in a
particular basis in terms of a large matrix in which the matrix
elements have rapidly varying phases and a smooth magnitude
distribution.  This problem, however, is too complicated to solve
exactly.  For an ideal system, the matrix is too large, and for a
realistic system, the matrix elements are unknown.  There are two
motivations for using a random matrix model: 1) In
the limit of small level spacings compared to energy scales of
interest, spectra are often insensitive to the 
details of matrix elements with uncorrelated phases
and the same average magnitude.  The robustness of the Wigner
semicircle distribution is an example.  2) A simple set of integral
equations may be derived for the spectrum.  These equations may be
solved numerically and compared directly with experiments.

The microscopic model for a superconductor in a magnetic field was
developed as a generalization of the zero-field BCS \cite{BCS} theory 
by Gor'kov \cite{gorkov59}, using a Green's function description,
and by de Gennes \cite{deGennes}, using a wavefunction description. 
The variational Hamiltonian is
\begin{equation}
\label{twobodyh}
 {\cal H}' = \int \!d{\bf r} 
\; \, \Psi_{\bf r}^\dagger \;
 \left[ \begin{array}{cc} {\cal H}_0({\bf r}) & \Phi({\bf r}) 
 \\ \Phi^\ast({\bf r})  & -{\cal H}_0^\ast({\bf r}) \end{array} \right]
\; \Psi_{\bf r}^{\vphantom{\dagger}} \ 
\end{equation}
where $\Psi_{\bf r}^\dagger \equiv \big[ \, c^\dagger_{{\bf r}\uparrow} \;
c^{\vphantom{\dagger}}_{{\bf r}\downarrow} \, \big]$, 
$c^\dagger_{{\bf r}\sigma}$ are the electron creation operators,
and
${\cal H}_0$ is the bare Hamiltonian ${\cal H}_0$ $\equiv$
$(1/2m)$ $\left( i{\bf\nabla} - e{\bf A}/c\right)^2-{E_F^{}} $.
The order parameter is
\begin{equation}
\label{selfcons}
 \Phi({\bf r}) = -V_0\,\Omega\,
 \langle c^{\vphantom{\dagger}}_{{\bf r}\uparrow} 
c^{\vphantom{\dagger}}_{{\bf r}\downarrow}  \rangle \ 
\end{equation}
for a system in a volume $\Omega$ and with a local two-body
interaction of strength $V_0>0$.  In the absence of a magnetic field,
plane waves diagonalize the bare Hamiltonian and the order parameter
may be taken to be a constant, $\Phi({\bf r})=\Delta_0$.
Eq. (\ref{twobodyh}) then separates into $2\times 2$ matrices,
yielding the BCS spectrum $E_{\bf k}=\sqrt{\varepsilon_{\bf k}^2 +
\Delta_0^2}$, where $\varepsilon_{\bf k} = \hbar^2 k^2/2m - E_F^{}$.

%% the separation between vortices is much less than a
%% penetration depth.  

In magnetic fields large compared to the field of first penetration,
$H_{c1}$, the field inside the superconductor is nearly uniform.
The eigenstates of ${\cal H}_0$, for an ideal system, are Landau
levels.  The momentum with respect to the vortex lattice, which
distinguishes states within a Landau level, is conserved, and
Eq. (\ref{twobodyh}) separates into $2N\times 2N$ matrices where $N
\sim \omega_D^{}/\omega_c$, the cutoff energy for the pairing
interaction over the Landau level spacing.  This formulation has been
used to calculate certain electronic properties of an ideal system
\cite{norman,bahcalla}.

A more general method \cite{bahcallb}, which does not depend on the
exact eigenstates being Landau levels, is to consider, as in the
Anderson description of dirty superconductors in no magnetic field
\cite{anderson}, arbitrary eigenstates $\psi_\alpha$ of the bare
Hamiltonian:
\begin{equation}
{\cal H}_0\; \psi_\alpha({\bf r}) \ = 
\ \varepsilon_\alpha\; \psi_\alpha({\bf r})\ .
\end{equation}
New quasiparticle operators may be defined with respect to this basis:
$d_{\alpha\sigma}^\dagger = \int\! d{\bf r}\; \psi_\alpha({\bf r})
c_{{\bf r}\sigma}^\dagger$.  The order parameter may be written
\begin{equation}
\Phi({\bf r}) \ \equiv\ \phi\; \chi({\bf r})\ ,
\end{equation}
where $\chi({\bf r})$ is normalized so that $\int |\chi({\bf
r})|^2\,d{\bf r} = \Omega$, and $\phi$, the spatial average, is real
and positive.  We can then define a pairing matrix
\begin{equation}
{\cal A}_{\alpha\beta} \ \equiv \ \int\!\!d{\bf r} \ \, \chi({\bf
r})\,\psi_\alpha^{\vphantom{\dagger}} ({\bf r})
\,\psi_\beta^{\vphantom{\dagger}}({\bf r}) \ ,
\end{equation}
so that the variational two-body Hamiltonian  is
\begin{equation}
\label{gentbh}
{\cal H}' =  \Psi^\dagger\;      \left[ \begin{array}{cc} \varepsilon
  & \phi\, {\cal A} \\
  \phi \,{\cal A}^\dagger & -\varepsilon \end{array}  \right]\,  \Psi\ ,
\end{equation}
where $\Psi$ is the vector of quasiparticle operators $\Psi^\dagger
\equiv \big[\, \cdots\, d_{\alpha\uparrow}^\dagger \cdots \; \cdots\,
d_{\beta\downarrow}^{\vphantom{\dagger}}\cdots \, \big]$ and
$\varepsilon$ is the diagonal matrix of bare eigenvalues
$|\varepsilon_\alpha|<\hbar\omega_D^{}$. The cutoff energy for the
pairing interaction, $\hbar \omega_D^{}$, is taken to be the
Debye energy in conventional superconductors.

In zero magnetic field, time-reversal invariance in an isotropic
superconductor ($\chi=1$) insures that the pairing matrix connects
only time reversed states: ${\cal A}_{\alpha\beta} \propto
\delta_{\alpha,\beta}$.  In the case of the ideal system, near
$H_{c2}$, it has been shown that ${\cal A}_{\alpha\beta} \propto
\exp(-|\varepsilon_\alpha -\varepsilon_\beta|^2/ W_0^2)$ where $W_0
\approx \Delta_0$, the zero-field gap \cite{norman,bahcalla}.  In
other words, the effect of the magnetic field is to ``fuzz'' out the
energy range over which states are paired, by an amount of order the
zero-field gap.  

The above pairing matrix description \cite{norman,bahcalla,bahcallb}
motivates the choice of the following model.  We consider an ensemble
of $2N\times 2N$ Hermitian matrices
\begin{equation}
\label{ensemble}
{H}\  =\  \left[ \begin{array}{cc} E_0
  & \phi \, { A} \\
  \phi {A}^\dagger & -E_0 \end{array}  \right]\ ,
\end{equation}
where $E_0$ is a diagonal matrix of $N$ uniformly distributed
eigenvalues $|\varepsilon_i|<1$, and the pairing matrix $A$ is
\begin{equation}
\label{ensadef}
A_{ij} \ \equiv \ {1\over\sqrt{N}}\; 
h(\varepsilon_i-\varepsilon_j)\ c_{ij} \ ,
\end{equation}
where $c_{ij}$ are selected from a complex Gaussian random
distribution $\langle c_{ij}^\ast c_{kl}^{}\rangle = \delta_{ik}
\delta_{jl}$.  The cutoff function $h$ is
\begin{equation}
h(x)\ \equiv \  {1\over \pi } \ {W\over x^2 \, + \, W^2}\ .
\label{ensfdef}
\end{equation}
The matrix $E_0$ models the spectrum of the normal metal in the range
$E_F^{}-\hbar\omega_D^{}<E< E_F^{}+\hbar\omega_D^{}$; all energies
have been normalized in units of the Debye energy.  The bare energy
level spacing $\delta=\hbar\omega_D^{}/N$ determines the size of the
matrix.  The level spacing is assumed to be small, so that the $N\to
\infty$ limit applies.  The cutoff function $h$ is related to the
Fourier transform of the time-dependence of the pair correlation
function.  The Lorentzian form used in Eq. (\ref{ensfdef}) is expected
on general grounds in a diffusive system (Ref. \cite{deGennes}, chap.\
8).  The spectrum is, however, not very sensitive to this choice of
cutoff function: a Gaussian cutoff yields results which differ by
$\alt 10\%$.

%% The diffusive model is reasonable for these systems, and yields a good
%% quantative fit with the data.

The two parameters of the model are $\phi$ and $W$.  $\phi$ is the
spatial average of the superconducting order parameter.  It decreases
from the gap value in zero field, $\phi(H=0)=\Delta_0$, to zero at the
phase transition, $\phi(H_{c2})=0$.  $W$ is the scale for the relative
energy difference between electrons being paired.  Time-reversal
symmetry gives $W(H=0)=0$; the scale increases steadily with magnetic
field, to $W(H_{c2})\equiv W_0$.  Solving the full self-consistent
equations for a given material would in principle determine the best
values for $\phi$ and $W$ as a function of applied field and
properties of that material.  This would also make the model much more
complex.  We therefore, instead, find solutions for arbitrary $\phi$
and $W$, and consider these as free parameters to be determined from
the data.

The spectrum of $H$ may be determined using methods similar to
those discussed in the context of solving large N 1+1 dimensional QCD
\cite{thooft} or 0+1 dimensional matrix-valued $\varphi^4$ theory
\cite{BIPZ}.  These techniques were further developed in the context
of a large N Anderson disorder model (many orbitals per lattice site)
by Wegner \cite{wegner} and in analyzing the eigenvalue correlators of
various matrix models by Brezin and Zee \cite{brezinzee}.

In general, the connection between random matrix models and 0+1
dimensional field theory follows from a Lagrangian of the form
$L=L_0+L'$, where the bare Lagrangian contains an $N$-vector of
fermions $\psi$ and an $N\!\times\! N$ scalar matrix $M$, $L_0 = i
\psi^\dagger \dot \psi + {\rm Tr}\, M^2$, and the interaction is $L' =
N^{-1/2}\psi^\dagger M \psi$. In the large N limit, the fermion
propagator for this theory is the ensemble-averaged Green's function
for $M$: $\int dt \, e^{iEt} \, \langle \psi^\dagger_i(t) \psi_j^{}(0)
\rangle = \langle (E-M)^{-1}_{\; ij}\rangle$.  This propagator may be
evaluated using standard Feynman diagram techniques.  The important
result is that in the large N limit only the planar (non-crossing;
generalized rainbow) diagrams survive
\cite{thooft,BIPZ,wegner,brezinzee}.

Some additional methods are necessary for the ensemble defined by
Eqs. (\ref{ensemble})--(\ref{ensfdef}) because of the Pauli matrix
structure and the cutoff function in $A$.  The Green's function for
$H$, ${\bf G}(E)\equiv 1/(E-H)$, may be written in the Gor'kov
notation,
\begin{equation}
    {\bf G} \ \equiv\  \left[ \begin{array}{cc} G  & F \\
    F' & G' \end{array}  \right]\ .
\end{equation}
Bold-faced quantities are $2\times 2$ matrices in electron-hole
space.  Dyson's equation is
\begin{equation}
    {\bf G} \ =\  {\bf G}_0 \ +\  {\bf G}_0 \; {\bf\Sigma} \; {\bf G}\ \ ,
\end{equation}
where ${\bf\Sigma}$ is the self-energy and ${\bf G}_0$ is the bare
Green's function ($\phi=0$).  In the large N limit, only the 
non-crossing diagrams survive, which implies
\begin{eqnarray*}
  {\bf\Sigma}_{ij} &=& \phi^2\ \sum_{k,l}\;
  \langle A_{ik}^{} A_{j l}^\ast \rangle
   \ \tau_1 \,{\bf G}_{kl}\, \tau_1 \\
  &=& \delta_{ij} \; {1\over N}\; \phi^2\ \sum_k \ \tau_1\,
      {\bf G}_{kk}\,\tau_1 \,\ h(\varepsilon_k-\varepsilon_i) \\
  &\equiv& \delta_{ij} \; \phi^2 \ \tau_1 \,
           {{\mbox{\boldmath${\rm S}$\unboldmath}}}(\varepsilon_i)\, \tau_1
\end{eqnarray*}
where $\tau_1$ is the Pauli matrix and we introduce the function
\begin{equation}
  {\mbox{\boldmath${\rm S}$\unboldmath}}(\varepsilon) \ \equiv \ 
  {1\over N} \ \sum_k\; {\bf G}_{kk}\; h(\varepsilon_k-\varepsilon) \ .
\label{introduces}
\end{equation}
Note that ${\mbox{\boldmath${\rm S}$\unboldmath}}$ depends on two
energies, $E$ and $\varepsilon$.  It is not a Green's function, it
is just a construct useful for solving this particular problem.
In the non-banded, uniform $A$ case we would have $h=1$ and
${\mbox{\boldmath${\rm S}$\unboldmath}}=(1/N) \,{\rm Tr}\, {\bf G}$,
independent of $\varepsilon$.

Inserting the result for ${\bf\Sigma}$ into Dyson's equation,
performing the sum in Eq. (\ref{introduces}), and replacing the
discrete bare energies $\varepsilon_i$ by the continuous variable
$\varepsilon$, yields
\begin{equation}
\label{matinteqn}
  {\mbox{\boldmath${\rm S}$\unboldmath}}(\nu)  = \int\limits_{-1}^1
\!\!  d\varepsilon\     {h(\nu-\varepsilon)\over 
  \left[{\bf G}_0^{-1}(\varepsilon)  - \phi^2 \; \tau_1 \,
  {{\mbox{\boldmath${\rm S}$\unboldmath}}}
  (\varepsilon) \, \tau_1   \right]} \ .
\end{equation}
This is an implicit integral equation for
${\mbox{\boldmath${\rm S}$\unboldmath}}$.  Writing
\begin{equation}
  {\mbox{\boldmath${\rm S}$\unboldmath}} \ \equiv \ 
  \left[ \begin{array}{cc} {\rm S}_1  & {\rm S}_3 \\
  {\rm S}_4 & {\rm S}_2 \end{array}  \right] \ ,
\end{equation}
and using that ${\bf G}_0^{-1}=E - \tau_3 \varepsilon$,
we have that the off-diagonal elements of the matrix in the
denominator of the integral equation are $S_3$ and $S_4$.  Hence, a
solution exists with $S_3=S_4=0$.  For a given theory, the spectrum in
the large N limit is unique, so we expect this to be the only
solution.  Eq. (\ref{matinteqn}) then reduces to
\begin{mathletters}
\label{solveme}
\begin{eqnarray}
  {\rm S}_1(\nu) &=& \int_{-1}^1 d\varepsilon\ \ {h(\nu-\varepsilon)
             \over E-\varepsilon-\phi^2\,{\rm S}_2(\varepsilon)}\\
  {\rm S}_2(\nu) &=& \int_{-1}^1 d\varepsilon\ \ {h(\nu-\varepsilon)
             \over E+\varepsilon-\phi^2\,{\rm S}_1(\varepsilon)}\ .
\end{eqnarray} 
\end{mathletters}
For a given $\phi$ and $W$, Eq. (\ref{solveme}) may be solved
numerically to obtain ${\rm S}_1(\varepsilon)$ and ${\rm
S}_2(\varepsilon)$ at any energy $E$.  The single-particle
excitation spectrum then follows from
\begin{equation}
   {\rm Tr}\; G(E) \ =\  \int_{-\infty}^{\infty} \! d\nu
   \ \, {\rm S}_1(\nu,E)
\label{spectrum}
\end{equation}
and the usual relation $\rho(E)=(-1/\pi)\, {\rm Im}\, {\rm Tr}\,
G(E+i\epsilon)$.

In the weak coupling limit, the bare BCS gap $\Delta_0 \ll \hbar
\omega_D^{}$, so both $\phi$ and $W$ in Eq.  (\ref{solveme}) are much
less than 1.  In that case, the limits of integration may be extended
to $\infty$ in both directions, $W$ may be scaled out, and solutions
depend only on the parameter $\phi/W$.

%% --------- TWO COLUMN STYLE : uncomment following lines
 \begin{figure}[b]
 \epsfysize=6cm\centerline{\epsfbox{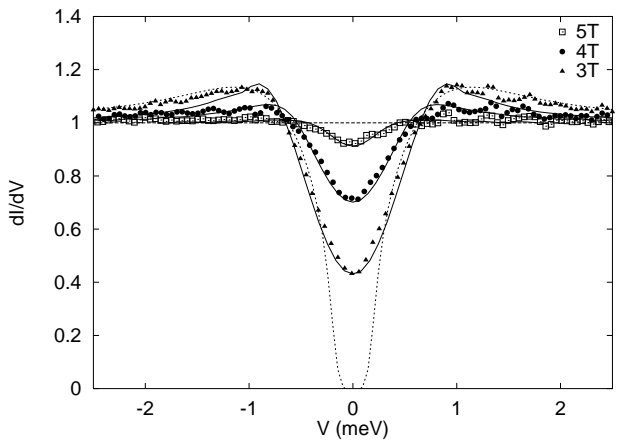}}
 \vspace{6pt}
 \caption{Normalized junction conductance for PbBi/Ge with
 $T_c=5.52\,\rm K$ at $360\,\rm mK$ for several magnetic fields [18].
 Solid lines: random matrix model ($\phi/W=.3$, $.8$, $1.6$).  Dashed
 line: Abrikosov-Gor'kov spectrum ($\zeta=.5$).}
 \label{figone}
 \end{figure}
%% ---------------------------

In Fig.\ 1 we compare the spectrum obtained this way with tunneling
spectra from planar tunnel junctions in BiPb/Ge thin films
\cite{hsuvalles}.  These spectra where taken at temperatures low
enough ($T=360\;\rm mK$) that thermal smearing affects the lineshape
by less than a few percent.  This thermal correction is included with
the standard expression \cite{tinkham} for the tunneling current
$dI/dV= -(G_0/\rho_0) \int\!  dE\, \rho(E)\, \partial f(E+eV)/\partial
V$, where the Fermi factor is $f(x)=1/(e^{\beta x}+1)$.  

The numerical agreement between the random matrix model and the
tunneling data is fairly close, a little worse at lower fields.  Also
shown in Fig.\ 1 is a fit with an Abrikosov-Gor'kov spectrum to the
$3\,\rm T$ data, in which the gap and pair-breaking strength are taken
as free parameters.  A pair-breaking strength \cite{maki} $\zeta=.5$
was used, which correctly reproduces the peak; a pair-breaking
strength $\zeta\approx 1$ gives more low energy states but fits poorly
at all energies.

One interesting feature of the random matrix model, as seen in Fig.\
1, is the presence of more low energy states at low fields than is
conventionally assumed.  The conventional view, however, has been
based on several indirect arguments rather than a direct solution of a
microscopic model.

First, the Abrikosov-Gor'kov solution for the Green's functions of a
superconductor in the presence of a dilute gas of magnetic impurities
\cite{abrigork} was applied to type-II superconductors by Maki, with
impurity concentration mapping to magnetic field strength \cite{maki}.
This mapping assumes, in addition to the dirty limit and local
electrodynamics, translationally invariant Green's functions (e.g.,
spatially uniform order parameter), which can not apply to a bulk
type-II superconductor at any magnetic field above $H_{c1}$.  Further,
it requires that the impurity averaging technique, valid for a dilute
gas of weak, uncorrelated, scatterers, apply to the case of a nearly
uniformly penetrating magnetic field distribution, which seems
unlikely.

Second, a vortex solution in the Ginzburg-Landau model has a small
``core'' of size $\xi$, the coherence length, compared to $\lambda$,
the penetration depth.  Although the Ginzburg-Landau model does not
contain electrons, the assumption is that the small core size over
which the order parameter magnitude is reduced acts as a box in which
electronic states are confined, leading to a small ``normal fraction''
of bound states.  However, the order parameter does not act on
electrons like a potential in a single-particle Schroedinger equation.
The order parameter enters the Bogoliubov-de Gennes equations
analogous to a spatially varying mass in a Dirac equation.  Squaring
the equation shows that currents also act as a confining potential.
A more direct analogy might be made with a tornado: it is not just
objects in the eye of the storm which are trapped and move with the
tornado, but also objects in the circulating currents around the eye
of the storm.

Bound states in the currents could explain why both the data and the
random matrix model indicate the presence of more low energy states
than expected from a Ginzburg-Landau picture.  Solutions for the
electronic structure around an isolated vortex would help address this
question; initial numerical work shows clear deviation from
Ginzburg-Landau behavior \cite{gygi}.  Note that if the bound states
can be shown to extend out to a penetration depth, then the overlap
with neighboring vortices, for fields above $H_{c1}$, would cause a
significant bandwidth.  This would further increase the expected low
energy density of states.

Experimentally, a surplus of low energy states is seen in the planar
tunnel junction measurements \cite{hsuvalles} as well as in local STM
measurements in the vortex state \cite{hess} and near SN junctions
\cite{tessmer}.  Deviations from the linear in magnetic field heat
capacity of the Ginzburg-Landau picture have been seen clearly in
high-T$_c$ superconductors \cite{moler}, as well as conventional
superconductors \cite{ramirez}.  More systematic studies of the
spectrum would help clarify the issue.

Another interesting feature of Eq. (\ref{solveme}) is that there is a
simple solution in the limit $W\ll\phi$.  Physically, this corresponds
to the situation where the energy scale for time-reversal symmetry
breaking is small, yet many bare eigenstates are mixed chaotically.
This might apply near SN junctions, in strongly anisotropic
superconductors, or at magnetic fields much smaller than $H_{c2}$.
(Note that it is necessary that many bare eigenstates be mixed, even
if $W$ is small, so that the large $N$ limit for the matrix ensemble
applies.)  In this small-$W$ limit, the cutoff function $h(x)$ becomes
a delta-function, and Eq. (\ref{solveme}) is solved by:
\begin{equation}
{\hat {\rm S}}_1({\hat E}, {\hat\varepsilon})
\ =\ ({\hat E}+{\hat\varepsilon}) 
\left[\ 1 - \sqrt{1-({\hat E}^2-{\hat\varepsilon}^2)^{-1}}\ \right]\ ,
\end{equation}
where the solution for $\sigma_2$ is given by $\sigma_2(E,\varepsilon) =
\sigma_1(E,-\varepsilon)$ 
and all energies scale with $\phi$: ${\hat E}=E/2\phi$,
${\hat\varepsilon}=\varepsilon/2\phi$, 
${\hat {\rm S}}_i=(2\phi)\sigma_i$.  The resulting spectrum
vanishes linearly near zero energy.  This offers a possible
alternative to the $d$-wave pairing explanation for the cuprate
superconductors, for which many observations have indicated close to a
linearly vanishing spectrum \cite{scalapino}.

We also note that mesoscopic superconducting dots have recently been
fabricated \cite{black}.  It is an interesting possibility that the
matrix ensemble discussed here might describe the eigenvalue
fluctuations of such systems, just as the standard GOE ensemble
describes the eigenvalue fluctuations of small metallic particles.

I would like to thank S.-Y.~Hsu and J.~M.~Valles for allowing me to
use their data, and D.-H.~Lee, D.~Shlyakhtenko, and D.~Voiculescu for
helpful discussions.  This work was supported by a Miller Research
Fellowship from the Miller Institute for Basic Research in Science.

%% ------------------------------ REFERENCES

%% ------ PREPRINT STYLE :   uncomment following lines
%% \newpage
%% \begin{figure}[b]
%% \epsfysize=6cm\centerline{\epsfbox{fig1b.ps}}
%% \vspace{6pt}
%% \caption{Normalized junction conductance for PbBi/Ge with
%% $T_c=5.52\,\rm K$ at $360\,\rm mK$ for several magnetic fields [18].
%% Solid lines: random matrix model ($\phi/W=.3$, $.8$, $1.6$).  Dashed
%% line: Abrikosov-Gor'kov spectrum ($\zeta=.5$).}
%% \label{figone}
%% \end{figure}
%% --------------------------

\end{document}